\documentclass{nature_Bian}

\usepackage{amssymb} 
\usepackage[implicit=false]{hyperref}
\usepackage{bbm}
\usepackage{mathrsfs}
\usepackage{amsfonts}
\usepackage{graphicx}
\usepackage{cite}
\usepackage[linesnumbered,ruled,lined]{algorithm2e}
\usepackage{array}
\usepackage{subfigure}
\usepackage{multirow}
\usepackage{color}
\usepackage{amsmath}
\usepackage{bm}
\usepackage{upgreek}
\usepackage{float}

\usepackage{cite}
\usepackage{authblk}
\usepackage{gensymb}

\title{Image-free single-pixel segmentation}

\author[]{Haiyan Liu,$^1$ Liheng Bian,$^{1,*}$ Jun Zhang,$^1$ }

\begin{document}

\maketitle

\begin{affiliations}
 \item School of Information and Electronics $\&$ Advanced Research Institute of Multidisciplinary Science, Beijing Institute of Technology, Beijing, 100081, China\\
 $^*$bian@bit.edu.cn
\end{affiliations}


\begin{abstract}
The existing segmentation techniques require high-fidelity images as input to perform semantic segmentation. Since the segmentation results contain most of edge information that is much less than the acquired images, the throughput gap leads to both hardware and software waste. In this letter, we report an image-free single-pixel segmentation technique. The technique combines structured illumination and single-pixel detection together, to efficiently samples and multiplexes scene's segmentation information into compressed one-dimensional measurements. The illumination patterns are optimized together with the subsequent reconstruction neural network, which directly infers segmentation maps from the single-pixel measurements. The end-to-end encoding-and-decoding learning framework enables optimized illumination with corresponding network, which provides both high acquisition and segmentation efficiency. Both simulation and experimental results validate that accurate segmentation can be achieved using two-order-of-magnitude less input data. When the sampling ratio is 1\%, the Dice coefficient reaches above 80\% and the pixel accuracy reaches above 96\%. We envision that this image-free segmentation technique can be widely applied in various resource-limited platforms such as UAV and unmanned vehicle that require real-time sensing.
\end{abstract}

\newpage

By classifying scene contents and subdividing them into different components or objects, scene segmentation extracts the target regions of interests to perform semantic analysis \cite{zaitoun2015survey,kuruvilla2016review,ghosh2019understanding}. It has found important applications in various fields. In medical diagnosis, segmenting organs and other substructures allows quantitative analysis of clinical parameters related to volume and shape \cite{litjens2017survey}. In intelligent transportation, real-time road segmentation is essential for autonomous driving \cite{badrinarayanan2017segnet}. In video surveillance, target segmentation is important for pedestrians detection of crowded real-world scenes \cite{leibe2005pedestrian}.

As computer vision develops in recent years, there have been various segmentation techniques based on either the conventional optimization-related machine learning \cite{kang2009comparative} or the recent network-related deep learning \cite{minaee2021image}. Despite of different frameworks, the existing segmentation techniques rely on extracting semantic features from high-fidelity input images \cite{song2017image}. In other words, these techniques require high-cost imaging systems to first acquire high-fidelity target images, and then implements semantic analysis to produce high segmentation accuracy. However, the information required for segmentation is mostly concentrated on component edges rather than the entire view in practice. In this sense, the acquired high-fidelity images generally contain a large amount of redundant data (such as target composition details and informationless background). The acquisition, transmission, storage and processing of these data waste both hardware (fine optics, transmission bandwidth, storage memory) and software (computing capabilities) resources and costs.

Recently, the image-free computational sensing concept has been proposed. Without the complex imaging process, image-free semantic inference enables action recognition and face recognition \cite{kulkarni2015reconstruction,lohit2015reconstruction} at a high compression rate. Based on the convolutional neural network, the image-free learning technique has achieved higher accuracy on image recognition datasets such as MNIST, ImageNet and CIFAR-10 \cite{lohit2016direct,adler2016compressed,9274326}. Most recently, H. Fu \emph{et al.} reported single-pixel sensing that classifies targets directly from a small amount of coupled single-pixel measurements, without the conventional image acquisition and reconstruction process \cite{fu2020single, Zhong2020Image}. The above studies validate the effectiveness of image-free sensing that implements high-level visual perception tasks without image acquisition and reconstruction. Nevertheless, the existing studies focus on target classification and recognition tasks that output only a single semantic conclusion, whose information throughput is much less than that of target segmentation that outputs a semantic map.

\begin{figure}
	\centering
	\includegraphics[width=0.8\linewidth]{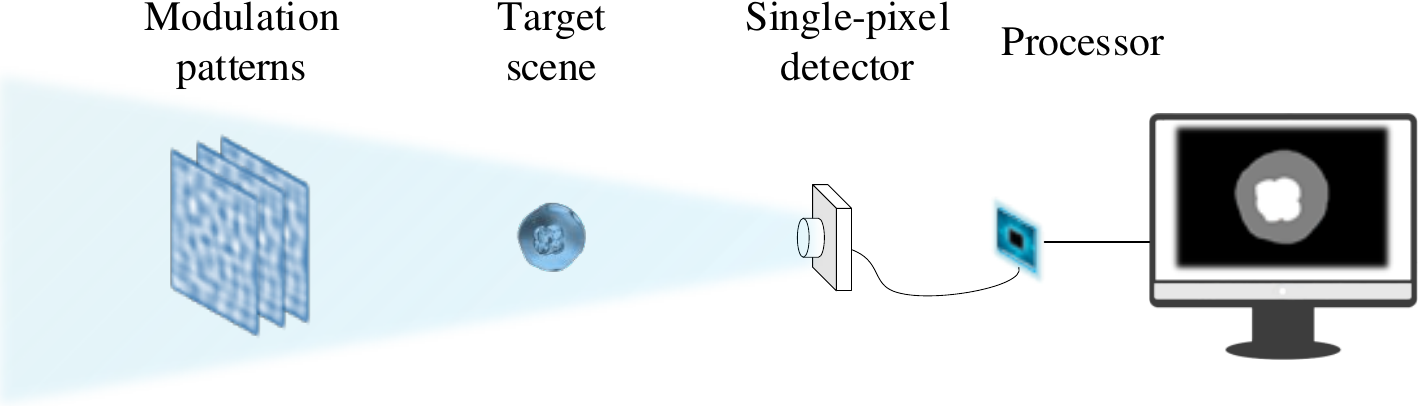}
	\caption{The reported image-free single-pixel segmentation framework.
		By acquiring a small amount of single-pixel measurements with optimized light modulation, the technique efficiently samples target's information and outputs segmentation results, without the conventional image acquisition and reconstruction process.}
	\label{fig:liu2021image-framework}
\end{figure}

In this letter, we report an image-free single-pixel segmentation technique that infers scene segmentation results from compressed non-visual measurements. Without the conventional image acquisition and reconstruction process, the approach directly extracts semantic features from one-dimensional single-pixel measurements. As shown in Fig. \ref{fig:liu2021image-framework}, the system shares a similar architecture to the single-pixel imaging system, which involves light modulation and single-pixel detection. Specifically, a spatial light modulator is employed to modulate illumination patterns, by which the target's feature information is compressed and encoded into one-dimensional intensity signals. A single-pixel detector is utilized to acquire the coupled measurements. Different from the conventional single-pixel imaging system, the single-pixel measurements are input into a neural network to directly output segmentation results without 2-D image reconstruction. Such an image-free framework effectively improves both acquisition and computational efficiency. Under a low sampling ratio, the reported technique enables high segmentation accuracy, and paves the way for light-load sensing where resources are expensive and limited.

\section*{Method}

\begin{figure*}
	\centering
	\includegraphics[width=\linewidth]{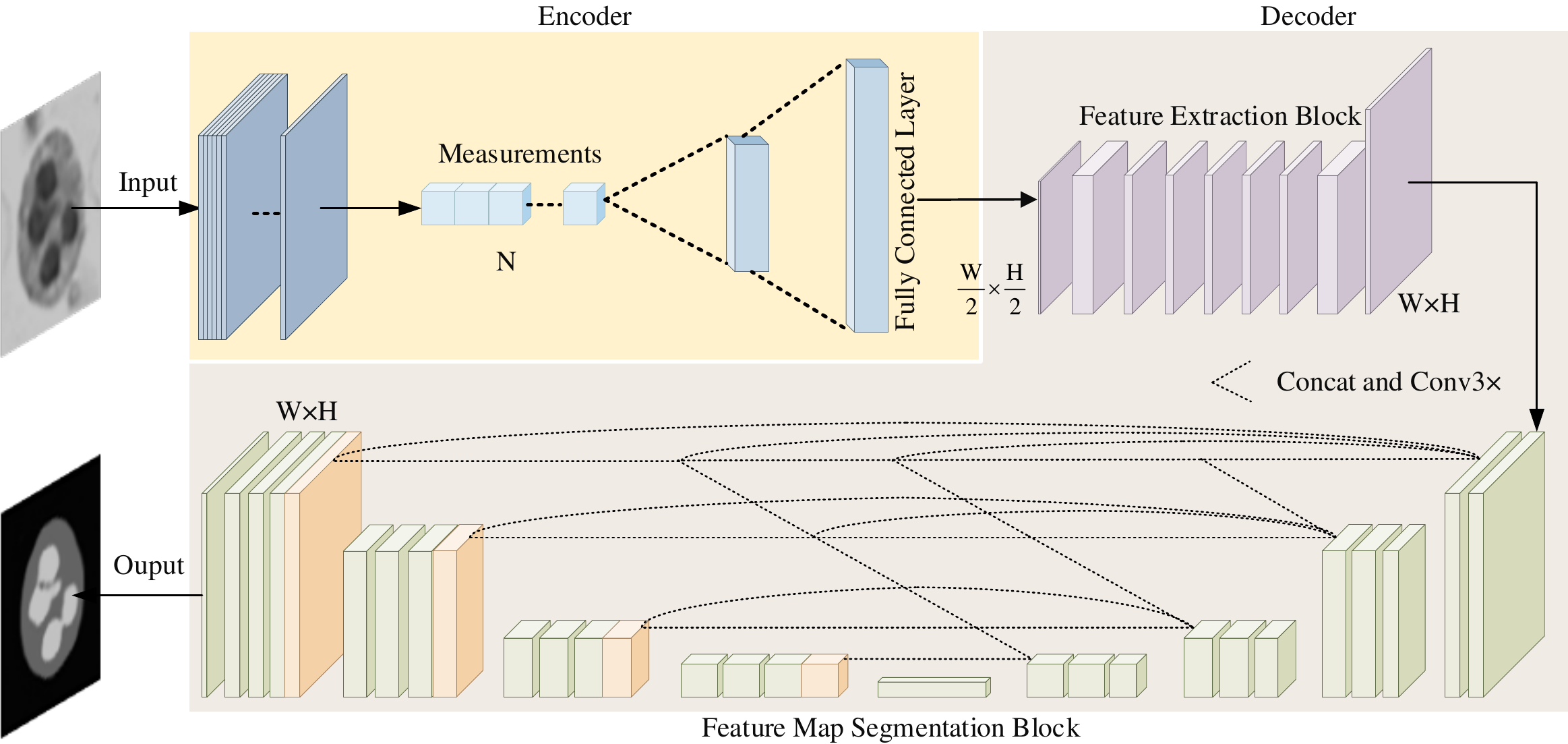}
	\caption{The single-pixel image-free segmentation network structure. The network consists of two parts, including an encoder module which encodes the target's information into one-dimensional measurements, and a decoder module that involves a feature extraction block and a feature map segmentation block. The decoder receives single-pixel measurements and directly outputs the segmentation results. The feature extraction block contains seven convolutional layers and one deconvolutional layer, and the size of the convolution kernels are $5\times 5$, $1\times 1$ and $9\times 9$, respectively. The feature map segmentation block mainly contains downsampling, upsampling and skip connection operations.}
	\label{fig:liu2021image-network}
\end{figure*}

\paragraph{The reported image-free single-pixel segmentation technique.} To further improve modulation and sensing efficiency, we built an end-to-end deep learning framework to simultaneously optimize the modulation patterns and the corresponding inferring network, as shown in Fig. \ref{fig:liu2021image-network}. It consists of two parts. The first part is an encoder module that modulates and couples the target light field into one-dimensional measurements. The second part is a decoder module that infers segmentation information from the non-visual measurements.

The encoder contains one convolution layer, in which the kernels represent the modulation patterns. The convolutional filters maintain the same size with the the target scene $I(x,y)$ ($64\times 64$ in this work). The $i$-th single-pixel measurement $S_{i}$ is mathematically modeled as
\begin{equation}
	S_{i}=\sum{I(x,y)\odot G_i(x,y)},
\end{equation}
where $G_i(x,y)$ represents the $i$-th modulation pattern, and $\odot$ denotes the Hadamard product. After the encoder outputs the one-dimensional measurements ($1\times 1\times N$), we added a full connected layer to extract more connotative semantic information. As a result, the data dimension was increased to $1\times 1\times 1024$, and then adjusted into a feature map of $1\times 32\times 32$ to input into the subsequent decoder.

With the input of a $32\times 32$ feature map produced by the encoder, the decoder outputs a $64\times 64$ segmentation map. The decoder is comprised of two parts. The first part is a feature extraction block designed based on the Fast Super-Resolution Convolutional Neural Network (FSRCNN) \cite{dong2016accelerating}, which includes five steps of feature extraction, shrinking, mapping, expanding and deconvolution. The feature extraction step utilizes 56 $5\times 5$ filters to first perform feature extraction on the input feature map. Then, the shrinking step adopts a smaller number of filters to reduce parameters and simplify the model, during which the feature dimension is reduced from 56 to 12. Next, the non-linear mapping part of 4 convolution layers of $3\times 3$ maps fuzzy features into clear features without changing feature map channels. Further, the expansion step adopts $1\times 1$ filters to extend the dimensions of the mapped high-resolution feature map and the number of channels back to 56, which is conducive to the reconstruction of a clearer segmentation map. Finally, the deconvolution step upsamples and aggregates the features using a set of deconvolution filters.

By applying the feature extraction block, a feature map with the size of  $64\times 64$ is obtained. Then, the second part of the encoder is a segmentation block designed based on the Unet++ architecture \cite{zhou2019unet++}. It mainly contains three steps of upsampling, downsampling and skip connection, which connects the feature maps of different scales. This segmentation block realizes feature extraction and fusion at different scales, and then outputs the final segmentation results.

Due to the lack of large-scale segmentation datasets, direct training of the reported network results in low segmentation accuracy (as validated in Fig. \ref{fig:liu2021image-trainstrategy}). Considering that large-scale natural image datasets are available, we derived a two-stage training strategy that intrinsically corresponds to the transfer learning theory \cite{tan2018survey}. In the first training stage, we apply large-scale natural image datasets to learn image-related prior features, by controlling that the gradient flow only goes back through the encoder and the feature extraction block. This helps better extract target features and improve sensing performance. In the second training stage, the gradient flow goes back through the entire network, namely that both the encoder and the decoder are updated at the same time. This operation transfers the learned features to the segmentation datasets, and optimizes the learning efficiency of the model. Once converged, the optimized filters of the encoder are set as light modulation patterns. Correspondingly, the single-pixel detector collects a sequence of coupled measurements that are input into the sensing decoder, which outputs the final segmentation results.

\section*{Simulations}

When training the network, we used the normalization initialization method with the bias initialized to 0, and used the Adam solver for gradient optimization. The weight decay was as 1e-4. The loss objective was set as the Mean Square Error. In the first training stage, we applied the STL-10 dataset \cite{coates2011stl10}. The learning rate was initialized as 2e-3, and was decreased by 0.8 for each 20 epochs. In the second training stage, the learning rate was set as 1e-3, and was decreased by 0.8 for each 50 epochs. We applied the White Blood Cell (WBC) segmentation dataset \cite{Zheng2018} which contains 300 $120\times 120$ images. The corresponding ground truth segmentation results were manually sketched by domain experts, where the nuclei, cytoplasms and background including red blood cells were marked in white, gray and black respectively. We expanded the dataset to 1200 images by horizontal and vertical mirroring, and then expanded these images three times through affine transformation and 50\degree rotation. Ultimately, we acquired 3600 images and randomly selected 2700 images to train the network, and evaluated its performance using the other 900 images. Each image is in gray scale and resized to $64\times 64$ pixels. The entire training process took $\sim$7 hours on a computer with an AMD 3700x processor and NVIDIA RTX 3090 graphics card.

\begin{table}
	\renewcommand\arraystretch{0.6}
	\caption{\bf Segmentation accuracy comparison of the two-stage and one-stage training strategies.}
	\centering
	\begin{tabular}{p{21mm}<{\centering}p{21mm}<{\centering}p{22mm}<{\centering}p{22mm}<{\centering}}
		
		\hline
		{Sampling}   &  \multirow{2}{*}{Metric}  &   {One-stage}  &   {Two-stage}   \\
		{ratio}  &  {}  &  {training}  &  {training}  \\		
		\hline
		
		\multirow{2}{*}{ 0.5 } &  PA  & 96.80  & \textbf{ 97.21 }\\
		{} &  DICE   & 80.88 & \textbf{81.72} \\
		\hline
		
		\multirow{2}{*}{ 0.05 } &  PA  & 94.35  &  \textbf{97.00} \\
		{} &  DICE   & 78.68 & \textbf{81.71} \\
		\hline
		
		\multirow{2}{*}{ 0.01 } &  PA  &  {77.90} &  \textbf{96.76} \\
		{} &  DICE   & {57.41} & \textbf{80.89} \\
		\hline
		
		\multirow{2}{*}{ 0.001 } &  PA  & 77.43  &  \textbf{94.40} \\
		{} &  DICE   & 57.12 & \textbf{78.15} \\
		\hline
		
		\multirow{2}{*}{ 0.0002 } &  PA  & 76.87  & \textbf{ 91.30} \\
		{} &  DICE   & 57.01 & \textbf{75.77} \\
		\hline
	\end{tabular}	
	\label{tab:liu2021image-table1}

\end{table}

To demonstrate the effectiveness of the reported two-stage training strategy, we first compared the segmentation accuracy of using the two-stage training strategy and the conventional one-stage training, as shown in Tab. \ref{tab:liu2021image-table1}. We employed two metrics to quantitatively evaluate segmentation accuracy, including the Pixel Accuracy (PA) and the Dice coefficient (DICE). The PA represents the percentage of correctly marked pixels, and the DICE is an ensembled measurement function to evaluate the structural similarity of two maps \cite{bertels2019optimizing}. The results in Tab. \ref{tab:liu2021image-table1} show that the two-stage training strategy obtains higher PA and DICE than the one-stage training at different sampling ratios. When the sampling ratio is 1\%, the Dice coefficient reaches 80.89\% and the pixel accuracy reaches 96.76\%. This validates that accurate segmentation can be achieved using two-order-of-magnitude less input data.

\begin{figure*}
	\centering
	\includegraphics[width=0.65\linewidth]{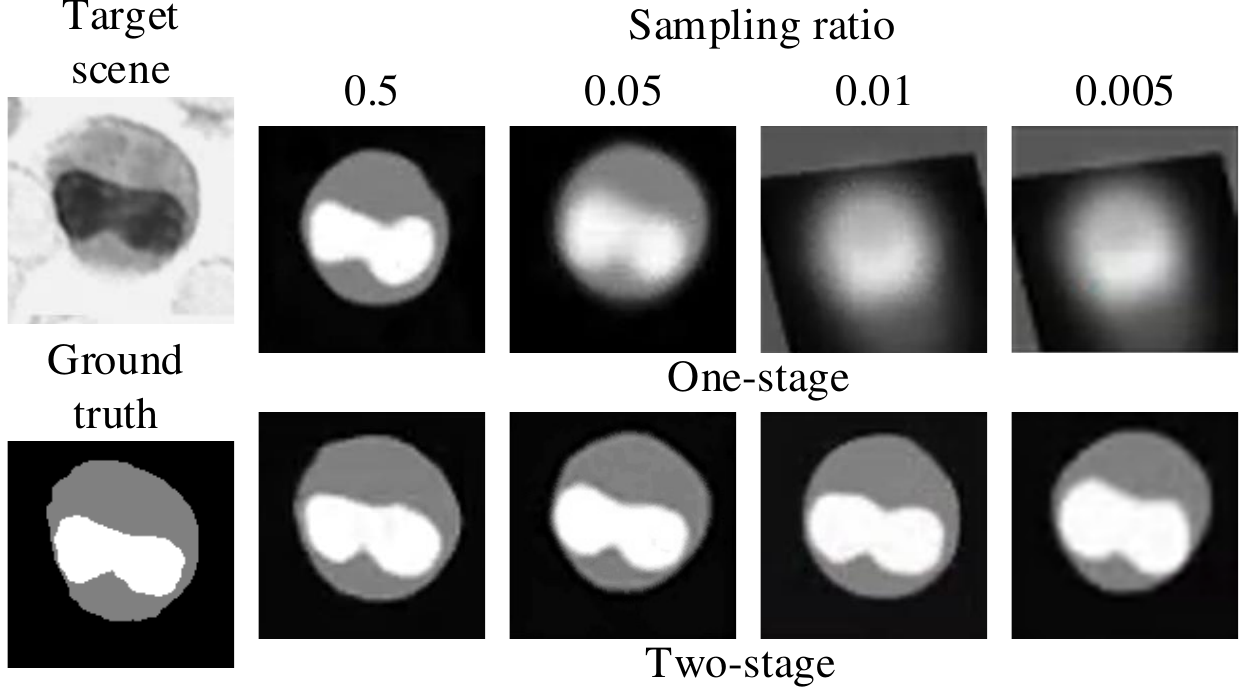}
	\caption{The comparison of the one-stage and two-stage training strategies at different sampling ratios on WBC dataset.}
	\label{fig:liu2021image-trainstrategy}
\end{figure*}

The exemplar segmentation maps of the two training strategies are presented in Fig. \ref{fig:liu2021image-trainstrategy}. We can see that as the sampling ratio reduces from 0.5 to 0.01, the segmentation results of the two-stage training strategy keep high fidelity compared to the ground truth. In contrast, the one-stage strategy produces distorted segmentation maps even at the sampling ratio of 0.5, which further degrade seriously as sampling ratio decreases. The superior performance of the reported two-stage training strategy originates from its intrinsic transfer learning nature. Specifically, the two-stage training runs between a source domain (classification features) and a target domain (segmentation features), which transfers image-based semantic knowledge to segmentation features of single-pixel measurements. Such a strategy effectively improves the generalization ability of the network and corresponding segmentation accuracy.

\begin{table}
	\renewcommand\arraystretch{0.6}
	\caption{\bf Segmentation accuracy on the WBC dataset under different modulation patterns (random patterns, Hadamard patterns and the optimized patterns) and different sampling ratios (0.0002-1).}
	\centering
	\begin{tabular}{p{15mm}<{\centering}p{15mm}<{\centering}p{20mm}<{\centering}p{20mm}<{\centering}p{20mm}<{\centering}}
		
		\hline
		{Sampling}   &  \multirow{2}{*}{Metric}  &  Random  &  Hadamard  &  Optimized \\
		{ratio}  &  {}  &  modulation  &  modulation   &  modulation  \\		
		\hline
		
		\multirow{2}{*}{ 0.1 } &  PA  & 96.28  &  96.95  & \textbf{ 97.08 }\\
		{} &  DICE   & 78.34 & 80.11 & \textbf{81.78} \\
		\hline
		
		\multirow{2}{*}{ 0.05 } &  PA  & 95.61  &  97.00  &  \textbf{97.01} \\
		{} &  DICE   & 78.49 & 80.84 & \textbf{81.71} \\
		\hline
		
		\multirow{2}{*}{ 0.01 } &  PA  & 94.39  &  96.51  & \textbf{ 96.76 }\\
		{} &  DICE   & 76.69& 79.61 & \textbf{80.89} \\
		\hline
		
		\multirow{2}{*}{ 0.0002 } &  PA  & 76.83  &  76.72  & \textbf{ 91.30} \\
		{} &  DICE   & 56.16 & 56.77 & \textbf{75.77} \\
		\hline
	\end{tabular}	
	\label{tab:liu2021image-table2}
\end{table}

Then, we trained another two networks with the encoder filters fixed to random and Hadamard modulation patterns \cite{zhang2017hadamard}, to illustrate the effectiveness of the reported optimized modulation. In the training process of these two networks, the encoder was fixed while only the decoder parameters were updated. The segmentation performance of different modulation strategies is presented in Tab. \ref{tab:liu2021image-table2}. The results show that using the optimized modulation obtains the highest accuracy (especially at low sampling ratios). Specifically, the segmentation accuracy of the optimized modulation at 0.01 sampling ratio is even higher than that of the other two modulation strategies at 0.1 sampling ratio. This validates that the optimized illumination patterns enable to better extract target features and improve segmentation accuracy.

Next, with the same set of single-pixel measurements as input, we compared the segmentation accuracy of the reported image-free technique with that of the conventional first-reconstruction-then-segmentation image-based methods at different sampling ratios. The two state-of-the-art single-pixel imaging reconstruction algorithms were introduced for comparison, including the total-variation-based reconstruction technique (TV-Rec) \cite{bian2018experimental} and the deep-learning-based reconstruction method (DL-Rec) \cite{higham2018deep}. The images reconstructed by these two techniques were input into the pre-trained UNET++ segmentation network to produce the final semantic maps.

\begin{figure*}
	\centering
	\includegraphics[width=0.75\linewidth]{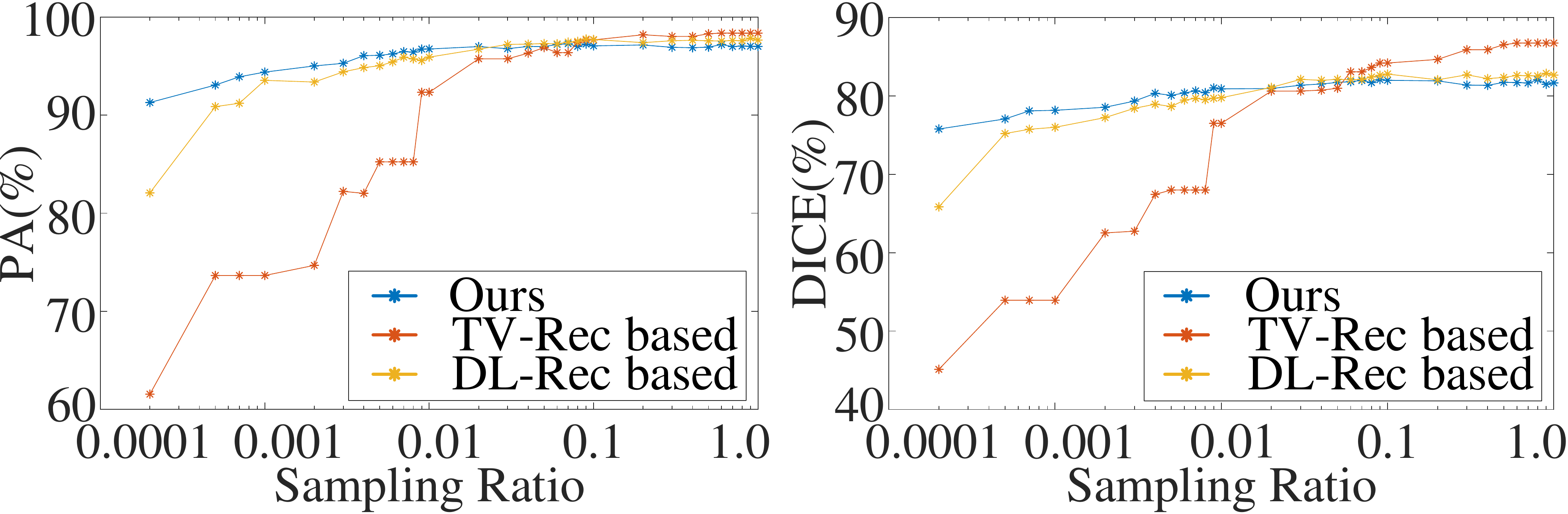}
	\caption{The comparison of segmentation accuracy between the reported image-free segmentation technique and the conventional first-reconstruction-then-segmentation method at different sampling ratios (0.0002-1). (a) and (b) show the PA and DICE evaluations respectively.}
	\label{fig:liu2021image-imaging}
\end{figure*}

\begin{figure*}
	\centering
	\includegraphics[width=0.65\linewidth]{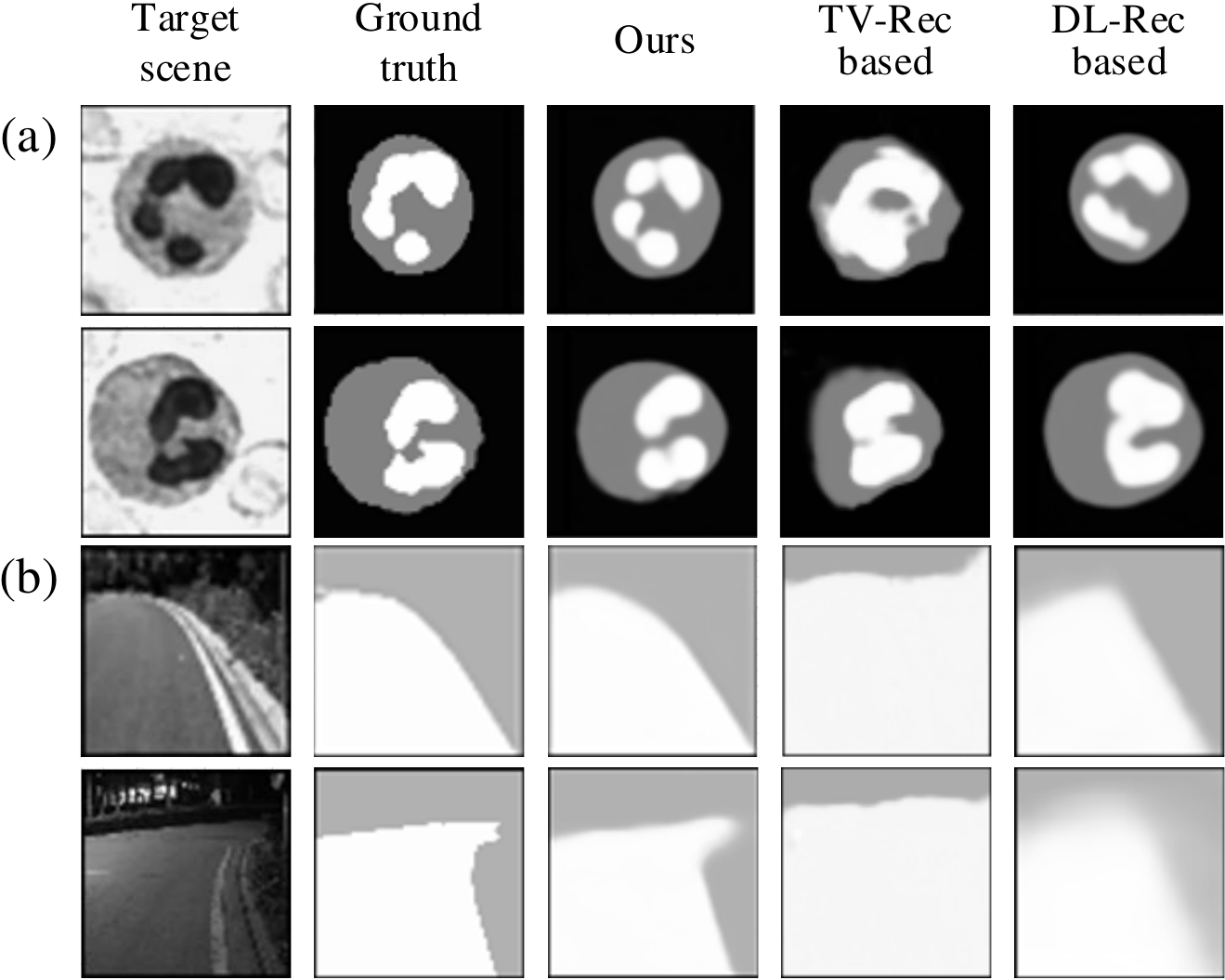}
	\caption{The segmentation maps on synthetic measurements, with the sampling ratio being 0.01. (a) and (b) show the results of the WBC dataset and the UAS dataset respectively.}
	\label{fig:liu2021image-imagingresult}
\end{figure*}

Figure \ref{fig:liu2021image-imaging} shows the quantitative comparison among the above different segmentation methods at different sampling ratios (0.0002-1). We can see that the TV-Rec technique enables to obtain higher segmentation accuracy when the sampling ratio is higher than 0.2, while its segmentation accuracy degrades quickly at low sampling ratios due to unsatisfied image reconstruction quality. The deep-learning reconstruction enables to improve reconstruction quality at low sampling ratios by a certain degree. In contrast, the reported image-free segmentation technique performs the best when the sampling ratio is lower than 0.02. As the sampling ratio reduces from 1.0 to 0.0002, the PA and DICE of the reported technique slightly decreases in the range of 6\%. Figure \ref{fig:liu2021image-imagingresult} (a) shows several exemplar segmentation maps of the three methods at 0.01 sampling ratio. We can see that the TV-Rec technique failed to segment both the nuclei and cytoplasms areas. Although the DL-Rec technique produces clear cytoplasms region compared with the TV-Rec method, there exist serious aberrations at its segmented nuclei regions compared with the ground truth. In contrast, the reported image-free segmentation technique produces the best visual quality among all the competing algorithms. It enables to distinctly segment nuclei and cytoplasms with less aberrations.

In addition, we also performed another simulation comparison of the different segmentation methods on the UESTC all-day Scenery (UAS) dataset \cite{zhang2018road}. The UAS dataset provides all-weather road images and corresponding binary labels, which discriminate the passable and impassable areas. We employed the images of four kinds of weather including sun, dusk, night, and rain, which contain 5670 training images and 710 testing images. We retrained and tested the image-free segmentation method and the conventional first-reconstruction-then-segmentation methods, and the results are presented in Fig. \ref{fig:liu2021image-imagingresult}(b) (the sampling ratio is 0.01). We can see that the image-free segmentation technique produces better segmentation results than the conventional methods. It produces clear passable areas with high quality at a low sampling ratio, while the TV-Rec technique failed to acquire road information to segment, and the DL-Rec technique produced wrong structure of passable areas.

\section*{Experiments}

\begin{figure*}
	\centering
	\includegraphics[width=0.55\linewidth]{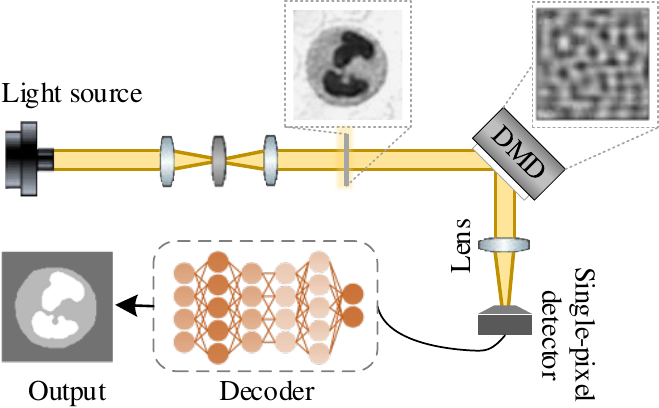}
	\caption{The proof-of-concept setup of image-free segmentation. The light source illuminated the film printed of the target scene. The DMD implemented the modulation of the pre-trained optimized patterns, and the single-pixel detector acquired the total intensity of the light field. The single-pixel measurements were input into the decoder network to output the final segmentation results.}
	\label{fig:liu2021image-setup}
\end{figure*}

\begin{figure*}
	\centering
	\includegraphics[width=0.65\linewidth]{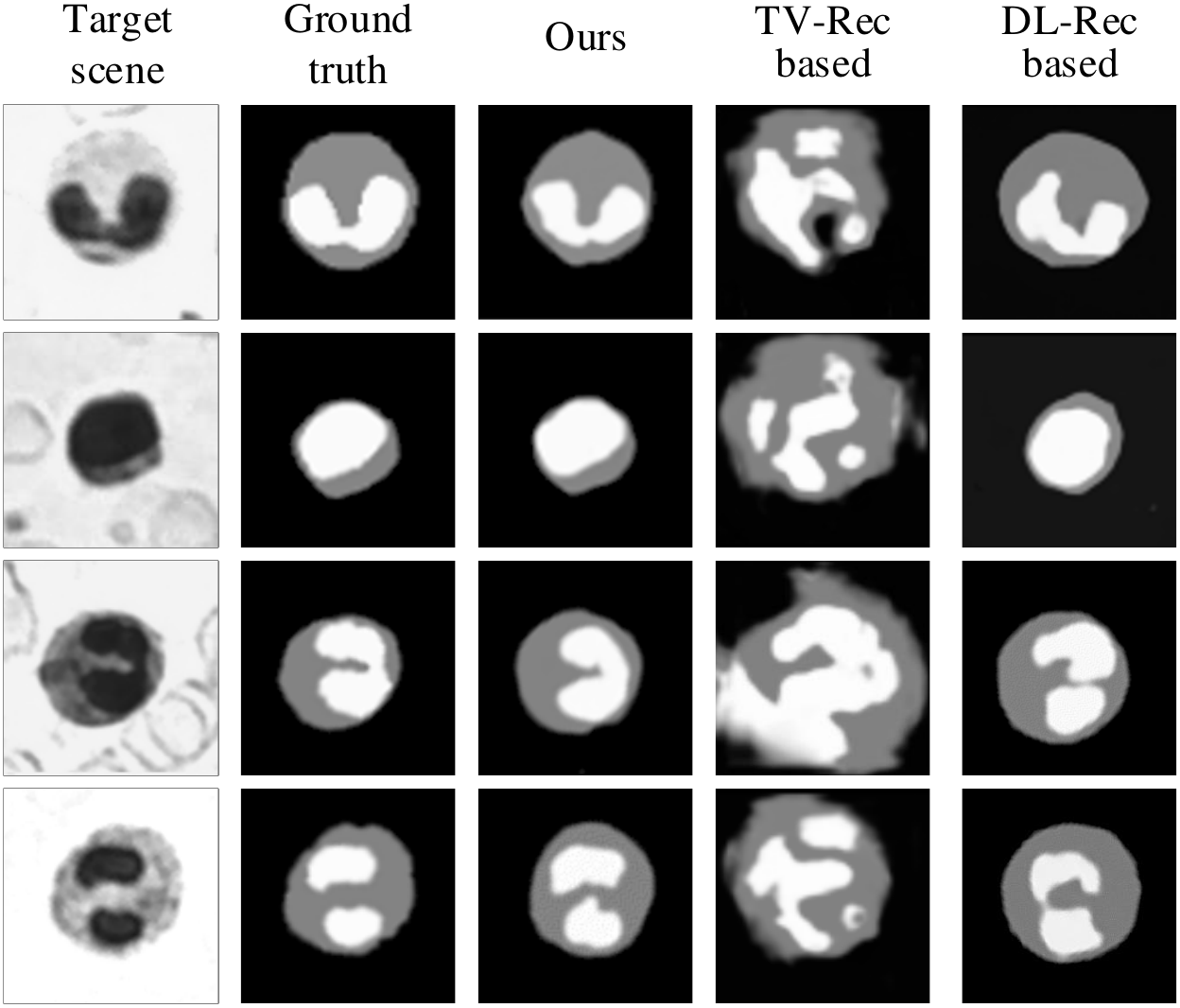}
	\caption{The segmentation results on experimental data.}
	\label{fig:liu2021image-experresult}
\end{figure*}

In order to further validate the effectiveness of the reported technique in practical applications, we built a proof-of-concept setup to acquire single-pixel measurements, as shown in Fig. \ref{fig:liu2021image-setup}. The pre-trained patterns were projected by a digital micromirror device (DMD, ViALUX V-7001) for illumination modulation. We set the sampling ratio as 0.01, and 40 patterns were projected for each target. The modulated light was projected onto a transmissive film printed of the target scene. The coupled signal was focused by a lens to an Si amplified photodetector (Thorlabs PDA100A2, 320?1100 nm). The measurements corresponding to different modulation patterns were input into the decoder network to output the final semantic segmentation results. The exemplar segmentation maps are presented in Fig. \ref{fig:liu2021image-experresult}. We can see that the nuclei and cytoplasms areas of the TV-Rec method were wrongly segmented, while the DL-Rec method maintains segmentation aberrations. In comparison, the results of the image-free segmentation technique are consistent with the ground truth with high fidelity.

\section*{Conclusion and Discussion}

In summary, we report a novel image-free single-pixel segmentation technique that maintains low hardware and software complexity. Different from the conventional first-reconstruction-then-segmentation methods that require 2-D image acquisition, transmission and processing, the reported technique directly produces segmentation maps from compressed 1-D measurements through an end-to-end neural network, which reduces both hardware and software complexity of the system. Besides, the modulation patterns in the single-pixel acquisition process are pre-trained and optimized together with the sensing network, which improves both the system's acquisition and segmentation efficiency. In such a computational sensing framework, the reported technique enables to effectively reduce data amount by two orders of magnitude, which might open up a new avenue for real-time scene segmentation on resource-limited platforms such as unmanned aerial vehicle and unmanned vehicle. 

The image-free single-pixel segmentation technique can be further extended. First, the segmentation accuracy can be further improved by introducing finely designed network modules such as the self-attention module \cite{dosovitskiy2020image}, which enables to focus on the areas that need to be segmented to improve segmentation accuracy. Second, although the learned modulation patterns achieve higher segmentation efficiency, they are at gray scale that consumes long implementation time using DMD or other light modulators compared to the binary patterns. We can further train binarized convolution kernels that maintain high implementation speed to lower acquisition time \cite{li2020discrete}. Third, the current network is running on the GPU server. To make the system applicable on resource-limited embedded platforms, we will further simplify the network model by judging the importance of different convolution kernels, and deleting filters with little loss of accuracy to reduce network scale \cite{molchanov2019importance,lin2020hrank}.

\vspace{5mm}




\begin{addendum}
 \item We thank Haiyan Liu, Liheng Bian and Jun Zhang for their valuable discussions and help. This work was supported by the National Natural Science Foundation of China (Nos. 61171119, 61120106003, and 61327902).
 \item[Author contributions] Liheng Bian and Haiyan Liu proposed the idea and designed the experiments. Haiyan Liu built the setup and conducted the experiments. All the authors contributed to writing and revising the manuscript, and convolved in discussions during the project.
 \item[Competing Interests] The authors declare no competing financial interests.
\end{addendum}

\newpage


\end{document}